\documentclass[final,5p,times,twocolumn]{elsarticle}

\usepackage{amsmath,amssymb}
\usepackage[utf8]{inputenc}
\usepackage{graphicx}
\usepackage{makecell}
\usepackage{mathrsfs}
\usepackage{bm}
\usepackage{bbm}
\usepackage{indentfirst}
\usepackage[table]{xcolor}
\usepackage{hyperref}
\usepackage[mathlines]{lineno}
\usepackage{booktabs}
\usepackage{braket}
\usepackage{placeins}
\usepackage{multirow}
\usepackage{slashed}
\usepackage{physics}
\usepackage{tabularx}

\hypersetup{
    colorlinks=true,
    linkcolor=blue,
    citecolor=blue,
    filecolor=blue,
    urlcolor=blue,
    linktoc=page
}
\journal{Physics Letters B}

\begin{document}

\begin{frontmatter}

\title{QCD chiral crossover line from Lee--Yang edge singularities}

\author[ccnu]{Heng-Tong Ding}

\author[bnl]{Swagato Mukherjee}

\author[bnl]{Peter Petreczky}

\author[ccnu]{Kai-Fan Ye\corref{cor1}}
\cortext[cor1]{Corresponding author.}
\ead{yekaifan@mails.ccnu.edu.cn}

\address[ccnu]{Key Laboratory of Quark and Lepton Physics (MOE) and Institute of
Particle Physics, Central China Normal University, Wuhan, 430079, China}

\address[bnl]{Physics Department, Brookhaven National Laboratory, Upton, New York, 11973, USA}

\begin{abstract}
We propose a universality-based reconstruction of the QCD chiral crossover line from Lee--Yang edge singularities in the complex baryon chemical potential plane.
The framework maps lattice-extracted complex Lee--Yang-zero estimates, treated as proxies for edge singularities, to the universal chiral Lee--Yang edge and thereby determines the $\mu_B$ dependence of both the chiral critical line in the light-quark chiral limit and the pseudo-critical crossover line at physical quark masses.
As an illustration, we apply the framework to Lee--Yang-zero estimates recently obtained by the Wuppertal--Budapest collaboration from high-statistics lattice QCD simulations.
Without imposing the previously determined small-$\mu_B$ expansion of the crossover line as input, the reconstructed curvature is consistent with existing continuum lattice-QCD results at small $\mu_B$.
The fitted chiral-limit transition temperature is also compatible with existing chiral-scaling analyses.
These results demonstrate that lattice information on Lee--Yang singularities, combined with universal chiral scaling, provides a quantitatively consistent constraint on the QCD crossover line within the present temperature window and establishes a framework that can be systematically improved with future Lee--Yang-zero determinations.
\end{abstract}


\end{frontmatter}

\section{Introduction}
The phase structure of quantum chromodynamics (QCD) at finite temperature and baryon density is a central topic in nuclear and particle physics. Of particular importance is the smooth but rapid crossover that separates the hadronic and quark-gluon plasma regimes. At zero baryon chemical potential $\mu_B$, lattice QCD has established the pseudo-critical temperature $T_{\rm pc}(0)\simeq 156.5\;\mathrm{MeV}$ with high precision~\cite{HotQCD:2018pds,Borsanyi:2020fev}. Determining how this crossover evolves with $\mu_B$ is essential for connecting first-principles theory with heavy-ion collision experiments and for understanding the thermodynamics of dense matter.

Extending lattice calculations to finite $\mu_B$ is, however, severely hindered by the sign problem. Present approaches therefore rely either on Taylor expansions around $\mu_B=0$ or on analytic continuation from simulations at imaginary $\mu_B$~\cite{Philipsen:2007rj,Ding:2015ona,Nagata:2021ugx}. Both strategies are intrinsically limited by the analytic structure of the QCD partition function in the complex chemical-potential plane: the radius of convergence of a Taylor series is bounded by the nearest non-analyticity, while a reliable analytic continuation requires control over the leading singularities. Consequently, existing lattice determinations of the crossover line are effectively restricted to $\mu_B/T\lesssim 2\text{--}3$ and to temperatures above $\sim 135\;\mathrm{MeV}$~\cite{HotQCD:2018pds,Borsanyi:2020fev,Borsanyi:2024xrx,Bonati:2018nut}.

A powerful framework for treating crossovers in a unified way is provided by the Lee--Yang edge singularities of the partition function~\cite{Yang:1952be,Lee:1952ig}. For a system that would undergo a true second-order transition in some limit, the nearest singularity in the complex plane of an external field (the Lee--Yang edge) controls the real-axis behavior. In QCD, the relevant external field for the chiral phase transition is the quark mass. In the chiral limit of light quark mass in $(2+1)$-flavor QCD, the transition is second order and the associated Lee--Yang edge lies on the real mass axis; for physical quark masses the edge moves into the complex plane and the transition on the real axis softens to a crossover, with the rapid variation still governed by the nearby singularity.

An analogous picture applies in the complex $\mu_B$ plane. If the crossover along the real $\mu_B$ axis is controlled by the chiral critical line, the nearest Lee--Yang edge should be tied to the chiral universality class. Extracting the location of this edge from lattice data therefore offers a direct route to reconstructing the physical crossover line, even when the edge itself lies well away from the real axis. Early work by Stephanov~\cite{Stephanov:2006dn} showed how universal scaling near the chiral limit determines the motion of singularities in the complex $\mu_B$ plane and how they constrain the radius of convergence of Taylor expansions. Subsequent studies have used lattice data at $\mu_B=0$ together with chiral critical scaling to estimate the Lee--Yang edge position and the associated convergence radius~\cite{Mukherjee:2019eou}. These developments motivate the strategy of using complex-plane
singularities to probe the QCD phase
structure~\cite{Dimopoulos:2021vrk,Basar:2023nkp,Wan:2024xeu,
Clarke:2024ugt,Wada:2024qsk,Wada:2026qzt}.
A remaining question is whether lattice-extracted
Lee--Yang-zero locations can quantitatively constrain the
nonuniversal chiral scaling map and thereby determine the
$\mu_B$ dependence of the physical crossover line, without
taking its known small-$\mu_B$ curvature coefficients as input.

Recent high-statistics lattice studies have begun to provide direct information on Lee--Yang zeros in the complex $\mu_B$ plane.
In particular, the Wuppertal--Budapest collaboration extracted Lee--Yang-zero estimates at several temperatures using rational approximations to the QCD free energy constructed from high-order baryon-number cumulants and imaginary-$\mu_B$ data~\cite{Adam:2025phc}.
Their analysis represents an important step toward using complex-plane singularities as quantitative lattice-QCD inputs.
It also shows that extrapolating the temperature dependence of Lee--Yang-zero locations to infer a critical endpoint can be highly sensitive to the chosen scaling or extrapolation ansatz.
In the present work we do not perform such a critical-endpoint extrapolation.

In this work we propose a reconstruction method that uses Lee--Yang edge locations as constraints on the nonuniversal map between QCD thermodynamic variables and the universal scaling variable of the chiral transition.
This differs from previous universality-based radius-of-convergence studies, where the chiral scaling map and nonuniversal parameters were used to infer the location of the Lee--Yang edge in the complex $\mu_B$ plane~\cite{Mukherjee:2019eou}.
Here the logic is reversed: the lattice-extracted Lee--Yang zeros are used to determine the QCD scaling map.
The central idea is to require that the complex singularities observed in the $\mu_B$ plane are mapped to the universal chiral Lee--Yang edge in the complex plane of the chiral scaling variable $z$.
Once this map is determined, it fixes the $\mu_B$ dependence of both the chiral critical line in the light-quark chiral limit and the pseudo-critical crossover line at physical quark masses.
The resulting crossover line is therefore reconstructed from singularities in the complex $\mu_B$ plane, without imposing the known small-$\mu_B$ curvature coefficients as input.

As an illustration, we apply the framework to the currently available Lee--Yang-zero estimates $\mu_{B,c}(T)$ reported in Ref.~\cite{Adam:2025phc}.
The reconstructed small-$\mu_B$ curvature is consistent with existing continuum lattice-QCD determinations, and the fitted chiral-limit transition temperature is compatible with existing chiral-scaling analyses.
The present study therefore provides both a consistency test of the chiral-scaling interpretation of lattice Lee--Yang singularities and a reconstruction framework that can be systematically improved as Lee--Yang-zero determinations become more precise.

\section{Method}
\label{sec:method}
We develop a universality-based framework to reconstruct the QCD
chiral crossover line from Lee--Yang edge singularities in the
complex baryon chemical potential plane. The central idea is to use
lattice-QCD determinations of complex singularities as constraints on
the nonuniversal mapping between QCD thermodynamic variables and the
universal scaling variable of the chiral phase transition.

Near the chiral limit, the singular part of the QCD free energy density can be written in the scaling form
\begin{equation}
f_s \sim H^{1+1/\delta}\Phi(z),
\qquad
H\equiv \frac{m_l}{m_s^{\rm phys}},
\label{eq:method_singularFreeEnergy}
\end{equation}
where $\Phi(z)$ is a universal scaling function and
\begin{equation}
z=z_0 H^{-1/(\beta\delta)}
\left[
\frac{T}{T_c(\mu_B)}-1
\right].
\label{eq:method_scalingVariable}
\end{equation}
Here $z_0$ is a nonuniversal normalization, $T_c(\mu_B)$ is the chiral critical line in the light-quark chiral limit, and $\beta$ and $\delta$ are the critical exponents of the three-dimensional $O(N)$ universality class~\cite{Karsch:2023rfb}.
For the numerical analysis, we specialize to the three-dimensional $O(2)$ universality class, appropriate for staggered fermions at finite lattice spacing~\cite{Ding:2024sux}.
The universal scaling function has a branch point at the Lee--Yang edge position
\begin{equation}
z_c=|z_c|\exp\left[\frac{i\pi}{2\beta\delta}\right].
\label{eq:method_universalEdge}
\end{equation}
Throughout the analysis we use $H=1/27$, $\beta\delta=1.6664(5)$, and $|z_c|=1.95(7)$~\cite{Karsch:2023rfb},
the latter being
consistent with functional renormalization group estimates~\cite{Connelly:2020gwa,Johnson:2022frg}.

We parameterize the chiral critical line as
\begin{equation}
T_c(\mu_B)=T_c^0\,g(x),
\qquad
x\equiv \frac{\mu_B}{T_c^0},
\label{eq:method_chiralLine}
\end{equation}
where $T_c^0\equiv T_c(0)$ is the chiral phase transition temperature at $\mu_B=0$ and the analytic mapping function satisfies $g(0)=1$.
For a Lee--Yang edge singularity located at $\mu_{B,c}$ at temperature $T$, the condition that it is mapped to the universal chiral Lee--Yang edge is
\begin{equation}
z_c
=
z_0 H^{-1/(\beta\delta)}
\left[
\frac{T}{T_c^0\,g(x_c)}-1
\right],
\qquad
x_c\equiv \frac{\mu_{B,c}}{T_c^0}.
\label{eq:edge_constraints}
\end{equation}
This complex equation is the central constraint of the reconstruction\footnote{The universal edge position $z_c$ is fixed by the branch-point non-analyticity of the singular scaling function $\Phi(z)$ and is not shifted by an additive regular
contribution. At the fixed value of $H$ used here, purely $H$-dependent
multiplicative corrections to the leading scaling fields may be absorbed
into the fitted $z_0$. Corrections with explicit $t$ dependence and
irrelevant-field corrections to scaling are neglected.}.
Given a set of Lee--Yang edge locations at different temperatures, one can determine the nonuniversal parameters entering $g(x)$, together with $T_c^0$ and $z_0$, by requiring all input singularities to map to the same universal edge $z_c$.

Within the leading-scaling approximation, a pseudo-critical line is
defined by a fixed real value $z=z_{\rm pc}$ of the scaling variable\footnote{The particular value of $z_{\rm pc}$---whether associated with
the peak of the chiral susceptibility or the inflection point of the
chiral condensate, or another pseudo-critical prescription---is irrelevant for the present reconstruction: any
fixed choice yields the same $\mu_B$ dependence and affects only the
normalization $T_{\rm pc}(0)$, as shown in the following two equations.}. Using Eqs.~\eqref{eq:method_scalingVariable} and~\eqref{eq:method_chiralLine}, one finds,
\begin{equation}
\begin{aligned}
T_{\rm pc}(\mu_B)
&=
T_c(\mu_B)
\left[
1+\frac{z_{\rm pc}}{z_0}H^{1/(\beta\delta)}
\right]
\\
&=
T_c^0
\left[
1+\frac{z_{\rm pc}}{z_0}H^{1/(\beta\delta)}
\right]
g\left(\frac{\mu_B}{T_c^0}\right).
\end{aligned}
\end{equation}
Identifying the $\mu_B$-independent prefactor with $T_{\rm pc}(0)$
then yields
\begin{equation}
T_{\rm pc}(\mu_B)
=
T_{\rm pc}(0)\,
g\left(\frac{\mu_B}{T_c^0}\right).
\label{eq:pc_line_physical}
\end{equation}
In this work, $T_{\rm pc}(0)$ is taken from continuum lattice-QCD determinations at zero baryon chemical potential.
Thus, within this approximation, the complex-plane information fixes the common leading $\mu_B$ dependence of the chiral critical line and the physical crossover line, while the overall normalization of the physical crossover line is set by $T_{\rm pc}(0)$.

The mapping function $g(x)$ describes the $\mu_B$ dependence of the
chiral critical line. Charge-conjugation symmetry requires it to be an
even function of $x$, while the normalization $T_c(0)=T_c^0$ implies
$g(0)=1$. We further require $g(x)$ to be analytic about $x=0$, real
and positive for real $x$, and non-increasing with $x^2$ on
the real-$\mu_B$ axis. To implement these properties, we write
\begin{equation}
g(x)=\exp[-F(w)],\qquad w=x^2,
\label{eq:method_mappingGeneral}
\end{equation}
with $F(0)=0$ and $F'(w)\geq 0$ for $w\geq 0$.

As a minimal flexible ansatz for describing the Lee--Yang-zero data, we
take $F(w)$ to be a quadratic polynomial in $w$,
\begin{equation}
{\rm M1:}\qquad
F_{\rm M1}(w)=a^2w+b^2w^2 ,
\label{eq:method_defaultMapping}
\end{equation}
and use this form for the reference analysis. To assess the dependence
of the reconstruction on the functional form of the mapping, we also
consider
\begin{align}
{\rm M2:}\qquad
F_{\rm M2}(w)
&=a^2w-abw^2+\frac{b^2}{3}w^3 ,
\label{eq:method_mapping2}
\\
{\rm M3:}\qquad
F_{\rm M3}(w)
&=\frac{a^2w+b^2w^2}{1+c^2w}.
\label{eq:method_mapping3}
\end{align}
Mappings M1 and M2 contain two parameters, $(a,b)$, whereas M3 contains
three, $(a,b,c)$. All three forms satisfy $g(0)=1$ and are non-increasing
functions of $x^2$ for real $x$, while M2 and M3 provide polynomial and rational
alternatives for estimating the mapping-ansatz dependence.

As discussed above a second order transition in QCD implies that there is a power-law nonanalyticity (branch cut) in the QCD free energy in complex $\mu_B$ that starts at Lee-Yang edge, $\mu_{B,c}$
at a given temperature in the infinite volume limit.
In finite volume numerical simulations this branch cut
becomes a series of zeros of the QCD partition function,
the so-called Lee-Yang zeros. The Lee-Yang zero closest
to the
origin
thus provides a good estimate of the
Lee-Yang edge, $\mu_{B,c}$ if the volume is sufficiently large.
The leading Lee-Yang zeros taken from Ref.~\cite{Adam:2025phc}  can be used as a proxy for Lee-Yang edge, $\mu_{B,c}$.

The mapping parameters, together with $T_c^0$ and $z_0$, are determined by requiring the edge locations predicted by a common scaling map to reproduce the lattice-extracted leading
Lee--Yang-zero locations within their uncertainties.

\section{Results}
\label{sec:results}

We apply the reconstruction method based on $O(N)$ scaling described in
Section~\ref{sec:method} to the leading Lee--Yang-zero estimates $\mu_{B,c}(T)$ reported in Ref.~\cite{Adam:2025phc}.
The $O(N)$ scaling is expected to be most reliable near the chiral phase transition temperature $T_c^0$.
The magnetic-equation-of-state analysis on $N_{\tau}=8$ lattices with the HISQ action gives $T_c^0=143.7(2)~\mathrm{MeV}$~\cite{Ding:2024sux}. We use this value as an external reference and leave $T_c^0$ free in our main fits. The results of Ref.~\cite{Ding:2024sux}  also indicate that the points at
$T=130$ and $155~\mathrm{MeV}$ may already lie outside the scaling window.
Our main analysis therefore uses the four temperatures
$T=135,140,145$, and $150~\mathrm{MeV}$.
Including the $T=130$ and/or $155~\mathrm{MeV}$ data in the analysis does not change
our conclusion qualitatively.

For each mapping function and Lee--Yang-zero input, the four
complex zero locations are fitted simultaneously in every bootstrap
realization. A common set of mapping parameters, together with common
values of $T_c^0$ and $z_0$, is therefore required to describe all four
temperatures. The universal $O(2)$ quantities specified in
Section~\ref{sec:method} are kept fixed. Unless stated otherwise, the quoted central
values and uncertainties denote the median and central $68.27\%$
interval of the bootstrap ensemble.
Details of the numerical analysis, including the study of the dependence on chosen temperature window are given
in \ref{app:numerical}.

We first investigate the dependence of the reconstruction on the functional
form of the mapping function. For this purpose, we use the Lee--Yang
zeros extracted from the pressure difference $\Delta P$ and perform
independent fits with mapping functions M1, M2, and M3.
For M1, M2, and M3, we obtain
$T_c^0=143.6^{+8.8}_{-14.8}$,
$144.2^{+10.3}_{-15.6}$, and
$141.8^{+10.2}_{-13.2}~\mathrm{MeV}$, respectively, with central values
spanning $141.8$--$144.2~\mathrm{MeV}$.
The corresponding fit qualities are $\chi^2/\mathrm{dof}=1.74^{+1.37}_{-0.91},\,
1.68^{+1.36}_{-0.88},\,
2.31^{+1.82}_{-1.21}$.
The three determinations of $T_c^0$ are mutually compatible and also consistent with the finite-cutoff $N_\tau=8$
magnetic-equation-of-state result,
$T_c^{N_\tau=8}=143.7(2)~\mathrm{MeV}$, reported in Ref.~\cite{Ding:2024sux}. The bootstrap distributions of $\chi^2/\mathrm{dof}$ do not reveal a clear preference among the mapping functions, although the median value is somewhat larger for M3.

For each bootstrap realization, the fitted mapping function is then
analytically continued to real $\mu_B$ and the physical crossover line is obtained
from~Eq.~\eqref{eq:pc_line_physical}.
For all reconstructions shown below, we adopt $T_{\rm pc}(0)=158.01(61)~\mathrm{MeV}$ from Ref.~\cite{Borsanyi:2020fev}. This input fixes the common
zero-density intercept of the reconstructed bands, whereas their $\mu_B$ dependence is determined
entirely by the fitted scaling map constrained by the Lee--Yang-zero inputs.

\begin{figure*}[!t]
\centering
\includegraphics[width=0.72\textwidth]
{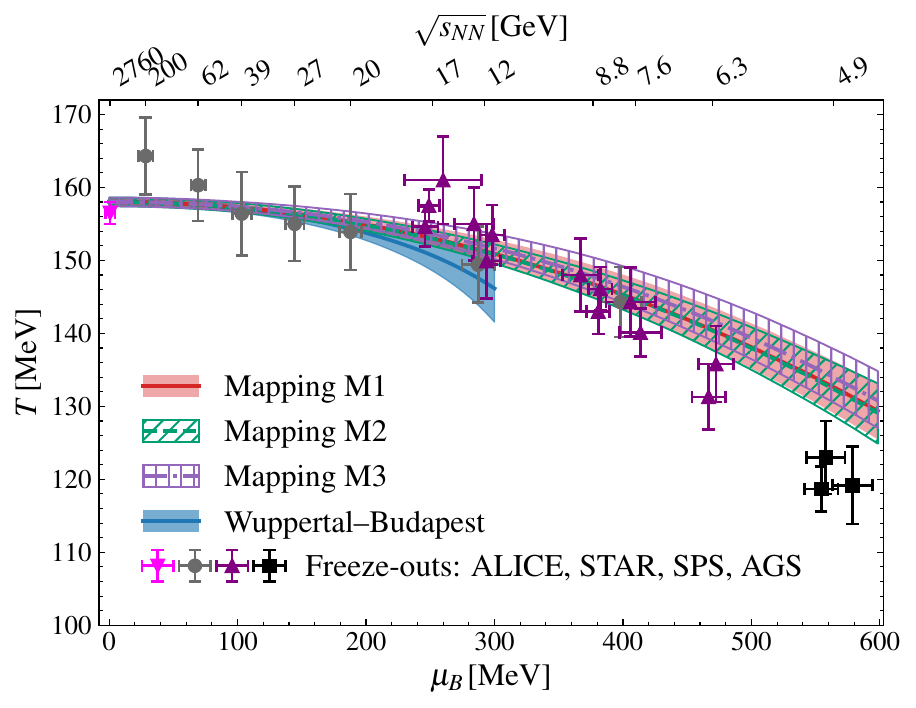}
\caption{
Dependence of the reconstructed QCD crossover line on the
mapping-function ansatz, using the Lee--Yang zeros extracted from
$\Delta P$.
The filled red band denotes Mapping~M1, while the hatched green and
purple bands denote M2 and M3, respectively.
The overlaid solid, dashed, and dash-dotted curves give the median M1,
M2, and M3 reconstructions, respectively, and each band shows the
central $68.27\%$ bootstrap interval.
The Wuppertal--Budapest crossover band
from Ref.~\cite{Borsanyi:2020fev} and phenomenological freeze-out estimates
are shown for comparison.
The freeze-out points are taken from Ref.~\cite{Andronic:2017pug} for
ALICE, Ref.~\cite{STAR:2017sal} for STAR, and Table~I of
Ref.~\cite{Cleymans:2005xv} for SPS and AGS.
Neither the real-axis lattice curvature nor the freeze-out estimates
are used as inputs to the reconstruction.
}
\label{fig:results_TpcMappings}
\end{figure*}

Fig.~\ref{fig:results_TpcMappings} compares the crossover lines
obtained with the three mapping functions M1, M2, and M3. The three bands are mutually
compatible within their bootstrap uncertainties over the displayed
range of $\mu_B$. Differences among the three parametrizations become
more visible toward larger $\mu_B$, where the analytic continuation is
less constrained, but remain smaller than the present statistical
uncertainties. The reconstruction is therefore not sensitive to
the functional form of the mapping within the current statistical accuracy.
To test whether the present input uncertainties conceal a stronger
dependence on the mapping, we repeated the bootstrap fits for the
four $\Delta P$ zeros with unchanged central values and both
semiaxes of each error ellipse reduced to $40\%$ of their original
lengths. At $\mu_B=600~\mathrm{MeV}$, the paired bootstrap
differences in $T_{\rm pc}$ relative to M1 are
$-0.46^{+0.48}_{-0.78}~\mathrm{MeV}$ for M2 and
$+2.67^{+0.65}_{-1.84}~\mathrm{MeV}$ for M3.
The mapping dependence thus remains at the level of a few MeV
in this sensitivity test.

At small and intermediate $\mu_B$, all three reconstructed bands are also
compatible with the existing continuum lattice-QCD crossover band from
Ref.~\cite{Borsanyi:2020fev}. This agreement is nontrivial because the
real-axis curvature is not used as an input in the reconstruction. The
reconstructed bands also broadly follow the systematics of the
phenomenological freeze-out estimates over much of the displayed
range. The freeze-out points, however, are not direct determinations of
the crossover temperature, and some of the lowest-energy points at the
largest $\mu_B$ lie below the reconstructed bands. They are therefore
included only as a phenomenological comparison.

Since the mapping-function dependence is mild, we
adopt~M1 for the remainder of the analysis. M1 is the
lowest-order analytic parametrization considered here. It contains two mapping
parameters, $a$ and $b$. M2
contains the same number of mapping parameters but
introduces a higher-order polynomial structure, whereas M3 contains
the additional parameter $c$.

\begin{figure*}[!t]
\centering
\includegraphics[width=0.72\textwidth]{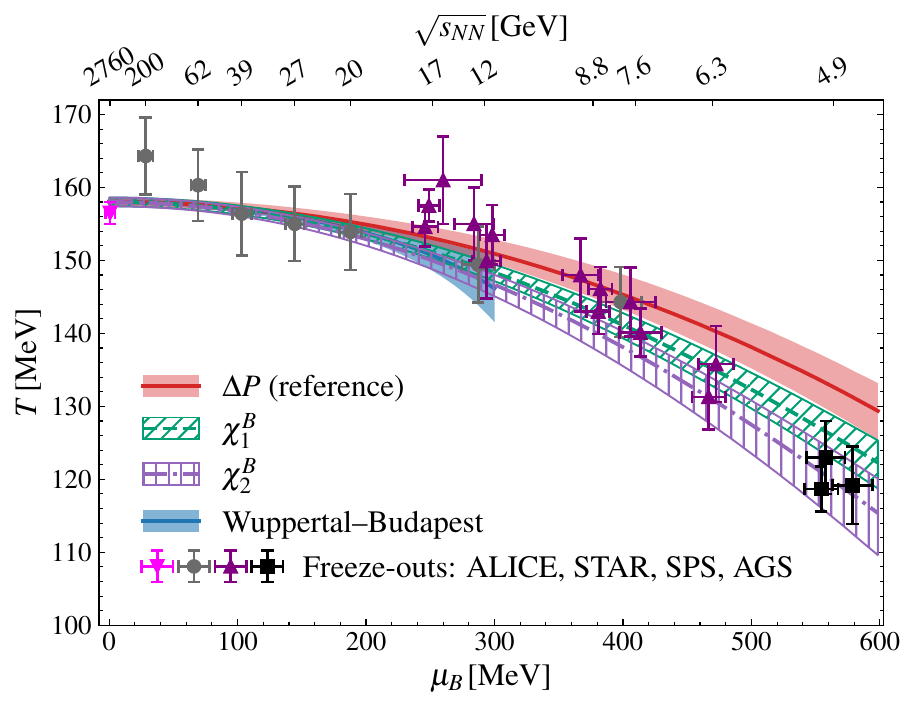}
\caption{Same as Fig.~\ref{fig:results_TpcMappings}, but for crossover lines
reconstructed with Mapping~M1 from the Lee--Yang zeros extracted from
$\Delta P$, $\chi_1^B$, and $\chi_2^B$.
The filled red band denotes the $\Delta P$ result, while the hatched
green and purple bands denote the $\chi_1^B$ and $\chi_2^B$ results,
respectively. The overlaid solid, dashed, and dash-dotted curves give
the corresponding median reconstructions.}
\label{fig:results_TpcReference}
\end{figure*}

We next repeat the M1 analysis using the leading Lee--Yang zeros
extracted from $\chi_1^B$ and $\chi_2^B$. The fits to the
$\chi_1^B$ and $\chi_2^B$ inputs give
$T_c^0=145.3^{+0.6}_{-4.7}~\mathrm{MeV}$ and $\chi^2/\mathrm{dof}=4.05^{+2.11}_{-1.63}$, and $T_c^0=136.8^{+2.7}_{-6.6}~\mathrm{MeV}$ and $\chi^2/\mathrm{dof}=2.99^{+1.86}_{-1.33}$, respectively. Both $T_c^0$ intervals overlap the broader interval obtained from the fit to $\Delta P$. The larger values of $\chi^2/\mathrm{dof}$ may indicate that the common scaling map provides a somewhat less satisfactory description of the
Lee--Yang-zero locations extracted from $\chi_1^B$ and $\chi_2^B$ than
of those extracted from $\Delta P$. We therefore regard these reconstructions primarily as cross-checks of the dependence on the observable used to determine the leading Lee--Yang zero. As shown in
Fig.~\ref{fig:results_TpcReference},
the reconstructions obtained from $\Delta P$, $\chi_1^B$, and $\chi_2^B$ are mutually compatible
at small $\mu_B$ and overlap the existing continuum
lattice-QCD crossover band.

At larger real $\mu_B$, the $\chi_1^B$
and $\chi_2^B$ bands decrease more rapidly than the $\Delta P$ band.
Their uncertainties also broaden, reflecting the increasing sensitivity of the continuation to the uncertainties of the complex zero locations. The spread among the bands thus provides an estimate of the present observable dependence of the reconstruction.

To quantify the small-$\mu_B$ behavior, for each input
$X\in\{\Delta P,\chi_1^B,\chi_2^B\}$ we expand the reconstructed
crossover line as
\begin{equation}
\frac{T_{\rm pc}^{X}(\mu_B)}{T_{\rm pc}(0)}
=
1-\kappa_2^{X}
\left(\frac{\mu_B}{T_{\rm pc}(0)}\right)^2
-\kappa_4^{X}
\left(\frac{\mu_B}{T_{\rm pc}(0)}\right)^4
-\kappa_6^{X}
\left(\frac{\mu_B}{T_{\rm pc}(0)}\right)^6
+\cdots.
\label{eq:results_kappaExpansion}
\end{equation}
Since Eq.~\eqref{eq:pc_line_physical} expresses the physical crossover line in terms of the fitted mapping function $g(\mu_B/T_c^0)$, the coefficients $\kappa_{2n}$ in Eq.~\eqref{eq:results_kappaExpansion} are derived from the fitted reconstruction parameters rather than fitted independently.
Eqs.~\eqref{eq:method_chiralLine} and
\eqref{eq:pc_line_physical} also imply that chiral critical and physical-mass crossover lines have the
same normalized small-$\mu_B$ dependence when expressed in terms of
$\mu_B/T_c^0$. Reexpressing the expansion in terms of
$\mu_B/T_{\rm pc}(0)$ introduces the normalization factor
$[T_{\rm pc}(0)/T_c^0]^{2n}$ in the coefficient of order $\mu_B^{2n}$.

For the $\Delta P$ reconstruction, we obtain
\begin{equation}
\begin{aligned}
\kappa_2^{\Delta P}
&=
0.012^{+0.002}_{-0.004},
\\
\kappa_4^{\Delta P}
&=
\left(0.33^{+2.64}_{-1.25}\right)\times10^{-4},
\\
\kappa_6^{\Delta P}
&=
\left(-9.0^{+13.1}_{-15.6}\right)\times10^{-7}.
\end{aligned}
\label{eq:results_kappaDeltaP}
\end{equation}
The corresponding results from the $\chi_1^B$ and $\chi_2^B$
reconstructions are
\begin{equation}
\begin{aligned}
\kappa_2^{\chi_1^B}&=0.017^{+0.002}_{-0.002},\\
\kappa_4^{\chi_1^B}&=\left(-1.45^{+0.89}_{-0.39}\right)\times10^{-4},\\
\kappa_6^{\chi_1^B}&=\left(8.2^{+3.6}_{-15.9}\right)\times10^{-7},\\[2pt]
\kappa_2^{\chi_2^B}&=0.020^{+0.003}_{-0.002},\\
\kappa_4^{\chi_2^B}&=\left(-1.42^{+2.45}_{-0.93}\right)\times10^{-4},\\
\kappa_6^{\chi_2^B}&=\left(2.9^{+13.6}_{-49.2}\right)\times10^{-7}.
\end{aligned}
\label{eq:results_kappaObservables}
\end{equation}

The leading curvature obtained from the $\Delta P$ input,
$\kappa_2^{\Delta P}=0.012^{+0.002}_{-0.004}$, agrees with the HotQCD
result $\kappa_2=0.016(6)$~\cite{HotQCD:2018pds} and the Wuppertal--Budapest result
$\kappa_2=0.0153(18)$~\cite{Borsanyi:2020fev} within uncertainties\footnote{The present reconstruction and the quoted HotQCD
coefficients correspond to $\mu_Q=\mu_S=0$, whereas the
Wuppertal--Budapest coefficients are obtained under strangeness neutrality, $n_S=0$, with $\mu_Q=0$~\cite{Borsanyi:2020fev}. HotQCD finds the results for $\mu_Q=\mu_S=0$ and for $n_S=0$, $n_Q=0.4n_B$ to be compatible within
uncertainties~\cite{HotQCD:2018pds}.}. The central values extracted from
the $\chi_1^B$ and $\chi_2^B$ inputs are larger, but show no statistically significant tension with these real-axis
determinations at the present precision. This comparison is particularly relevant because none of the existing real-axis curvature results is imposed as an input to
the Lee--Yang-zero fits.

The fourth-order coefficient from the $\Delta P$ reconstruction is
also compatible with the currently available lattice determinations, $\kappa_4=7.9(6.8)\times10^{-4}$ from the Wuppertal--Budapest collaboration~\cite{Borsanyi:2020fev} after conversion to the convention of
Eq.~\eqref{eq:results_kappaExpansion}, and $\kappa_4=10(70)\times10^{-4}$ from HotQCD~\cite{HotQCD:2018pds},
although it is much less precisely constrained than $\kappa_2$. The corresponding $\chi_1^B$ and $\chi_2^B$ results are likewise
compatible with these determinations at the present precision, while
their spread reflects the current observable dependence. The sixth-order coefficient is presently only weakly
constrained. Consequently, the leading coefficient $\kappa_2$ provides
the most robust quantitative small-$\mu_B$ test of the reconstruction.

The reconstructed bands should be interpreted as universality-based constraints obtained from the presently available estimates of finite-volume Lee--Yang zeros, which are used as proxies for thermodynamic-limit edge singularities. In particular, their continuation toward larger real
$\mu_B$ does not constitute a continuum determination of the QCD
crossover line; such a determination will require Lee--Yang-zero inputs
extrapolated to the thermodynamic and continuum limits.

\section{Conclusions}
We have introduced a universality-based reconstruction of the QCD chiral crossover line from Lee--Yang singularities in the complex baryon chemical-potential plane. The key step is to reverse the usual
use of chiral scaling: rather than employing a previously determined scaling map to predict the complex singularities, we use the
lattice-extracted leading Lee--Yang zeros to constrain the nonuniversal map between QCD thermodynamic variables and the universal scaling variable. Within leading scaling, the resulting map determines
the same normalized $\mu_B$ dependence of both the chiral critical line
in the light-quark chiral limit and the physical-mass
pseudo-critical line. The crossover line can therefore be reconstructed from complex-plane information without imposing its known real-axis
curvature as an input.

Using the leading Lee--Yang zeros extracted from $\Delta P$, the reference analysis gives $T_c^0=143.6^{+8.8}_{-14.8}~\mathrm{MeV}$, compatible with the existing finite-cutoff chiral-scaling determination. The reconstructed leading
curvature, $\kappa_2^{\Delta P}=0.012^{+0.002}_{-0.004}$, agrees with the HotQCD
and Wuppertal--Budapest continuum lattice-QCD determinations. This agreement provides an independent consistency test of the reconstruction, since the real-axis curvature is not used to determine the scaling map. The compatibility of the M1--M3 bands indicates only mild dependence
on the mapping-function ansatz at the present precision. Reconstructions
based on the $\chi_1^B$ and $\chi_2^B$ zeros provide additional checks
of the observable dependence, although their larger
$\chi^2/\mathrm{dof}$ values and their steeper decrease at large real
$\mu_B$ indicate that this dependence is not yet negligible. The
higher-order curvature coefficients remain considerably less well
constrained than $\kappa_2$.

The present analysis is a universality-based reconstruction from the currently available finite-volume Lee--Yang-zero inputs, with a continuum determination of the QCD crossover line left to future work.

Future Lee--Yang-zero determinations on larger spatial volumes and finer lattice spacings could enable more controlled assessments of finite-volume and discretization effects, including possible
contamination from unphysical branch points introduced by staggered
rooting~\cite{Giordano:2019gev}. More precise determinations with denser
temperature coverage near $T_c^0$ could further test the common scaling description and clarify its range of applicability. Such improvements would further establish Lee--Yang-edge information as an independent constraint on both the QCD chiral phase transition line and the physical-mass crossover line beyond the small-$\mu_B$ regime.

\section*{Declaration of competing interest}
The authors declare that they have no known competing financial interests or personal relationships that could have appeared to influence the work reported in this paper.

\section*{Acknowledgements}
We thank Nu Xu and the participants of the ECT* workshop ``Analytic structure of QCD and Yang-Lee edge singularity'' (Trento, 2025) for valuable discussions.

This material is based upon work supported by the U.S.~Department of Energy, Office of Science, Office of Nuclear Physics through Contract No.~DE-SC0012704 and within the frameworks of Scientific Discovery through Advanced Computing (SciDAC) award Fundamental Nuclear Physics at the Exascale and Beyond. This work is supported partly by the National Natural Science Foundation of China under Grant Nos.~12325508, 12293064, and 12293060 as well as the National Key Research and Development Program of China under Contract No.~2022YFA1604900 and the Fundamental Research Funds for the Central Universities, Central China Normal University under Grant Nos.~30101250314 and 30106250152.

The numerical analyses were carried out using the Nuclear Science Computing Center at Central China Normal University ($\mathrm{NSC}^{3}$).

\section*{Data availability}
Data will be made available on request.

\appendix
\section{}
\label{app:numerical}

We digitize the locations of Lee-Yang zeros and the corresponding error ellipses from Fig.~6 of Ref.~\cite{Adam:2025phc}.
The reference fit uses one complex zero at each of four temperatures, $T=135,140,145$ and 150 MeV.
This gives eight real data values. M1 and M2 have four fit
parameters, $(a,b,T_c^0,z_0)$. M3 has a fifth parameter, $c$.
We fit the zeros extracted from $\Delta P$, $\chi_1^B$, and $\chi_2^B$
separately. We retain the real--imaginary covariance.
Each ellipse's orientation and semi-axes define a Gaussian covariance
between the real and imaginary parts. We take the semi-axes as standard
deviations.
Each fit uses $10^4$ parametric-bootstrap realizations.
For each realization, we sample the Gaussian inputs and refit all
temperatures jointly. We also propagate uncertainties in the universal
inputs and $T_{\rm pc}(0)$.

Including data at $T=155~\mathrm{MeV}$, $T=130~\mathrm{MeV}$ or both leads to significantly larger $\chi^2$.
The fits with $T_c^0$ as a free parameter result in
values of the chiral transition temperature of about 120 MeV or smaller, i.e. in values that are even significantly smaller than the known value of $T_c^0=132^{+3}_{-6}~\mathrm{MeV}$
in the continuum limit~\cite{HotQCD:2019xnw} and $\kappa_2$ values which are also much smaller than the established one, suggesting
the above assertion that $T=130$ MeV and $155$ MeV are outside the scaling regime. If we perform fits with $T_c^0$
fixed to $143.7$ MeV we also see that the $\chi^2$ increases substantially if $T=130$ MeV and/or $T=155$ MeV
data are included. However,  the crossover
line shown in Figs~\ref{fig:results_TpcMappings} and~\ref{fig:results_TpcReference}, which is our
main result, does not show significant variation when $T=130$ MeV and/or $T=155$ MeV data are included.
For these reasons, we base our main results on fits restricted to
the temperature window $T=135$--$150~\mathrm{MeV}$.

\bibliographystyle{elsarticle-num}
\bibliography{ref}

\end{document}